\documentclass[fleqn,10pt]{wlscirep}
\usepackage[utf8]{inputenc}
\usepackage[T1]{fontenc}
\usepackage{bm}

\title{Toward Interpretable and Generalizable AI in Regulatory Genomics}

\author[1,*]{Masayuki Nagai}
\author[1,*]{Alan E. Murphy}
\author[1]{Kaeli Rizzo}
\author[1,$\dagger$]{Peter K. Koo}
\affil[1]{Simons Center for Quantitative Biology, Cold Spring Harbor Laboratory, Cold Spring Harbor, NY, USA}

\affil[*]{These authors contributed equally.}
\affil[$\dagger$]{e-mail: koo@cshl.edu}

\begin{abstract} 
Deciphering how DNA sequence encodes gene regulation remains a central challenge in biology. Advances in machine learning and functional genomics have enabled sequence-to-function (seq2func) models that predict molecular regulatory readouts directly from DNA sequence. These models are now widely used for variant effect prediction, mechanistic interpretation, and regulatory sequence design. Despite strong performance on held-out genomic regions, their ability to generalize across genetic variation and cellular contexts remains inconsistent. Here we examine how architectural choices, training data, and prediction tasks shape the behavior of seq2func models. We synthesize how interpretability methods and evaluation practices have probed learned \textit{cis}-regulatory organization and highlighted systematic failure modes, clarifying why strong predictive accuracy can fail to translate into robust regulatory understanding. We argue that progress will require reframing seq2func models as continually refined systems, in which targeted perturbation experiments, systematic evaluation, and iterative model updates are tightly coupled through AI–experiment feedback loops. Under this framework, seq2func models become self-improving tools that progressively deepen their mechanistic grounding and more reliably support biological discovery.
\end{abstract}

\begin{document}
\flushbottom
\maketitle

\thispagestyle{empty}

\section*{Introduction} 

Gene regulation is shaped by complex interactions among sequence features encoded in DNA that specify transcription factor (TF) binding, influence chromatin accessibility, and coordinate interactions between distal regulatory elements such as enhancers and promoters. These sequence-dependent mechanisms are commonly framed in terms of the \textit{cis}-regulatory code: the principles by which DNA sequence specifies regulatory activity in a given cellular context. Under this view, modeling gene regulation is centered around learning a quantitative mapping from DNA sequence to regulatory activity.

Recent advances in deep learning have improved our ability to predict regulatory activity directly from DNA sequence. Sequence-to-function (seq2func) models achieve strong performance across diverse experimental readouts of cell-type–specific regulatory activity, including chromatin accessibility, TF binding, histone modifications, transcription initiation, and steady-state gene expression \cite{avsec2021effective,chen2022sequence,dudnyk2024sequence,jaganathan2025predicting,cochran2024dissecting,he2024dissection,linder2025predicting,avsec2025alphagenome,barbadilla2025predicting,boshar2025foundational}. Here, \emph{function} refers to molecular regulatory activity measured by functional genomics assays, rather than cellular, organismal, or evolutionary processes. Seq2func models are now widely used for single-nucleotide variant (SNV) effect prediction \cite{sokolova2024deep}, hypothesis generation via model interpretation \cite{novakovsky2023obtaining}, and synthetic regulatory sequence design \cite{vaishnav2022evolution,taskiran2024cell,de2024targeted,gosai2024machine,de2025modelling}.

Despite this progress, key questions remain about how seq2func models generalize under perturbation and where their capabilities break down. Many models perform well on genomic sequences similar to their training data, yet their behavior under genetic perturbations or shifts in cellular context can be inconsistent \cite{karollus2023current,sasse2023benchmarking,kathail2024current}. These limitations reflect biased coverage in available training data, objectives used during training, and evaluation practices that incompletely probe out-of-distribution behavior.

In this Review, we survey the current landscape of genomic AI through the lens of seq2func modeling. We examine how architectural choices, training data, prediction tasks, model interpretation, and evaluation strategies shape seq2func model behavior and generalization. We outline a direction in which perturbation data, rigorous evaluation, and iterative model refinement become central to model development and assessment, enabling seq2func models to become more mechanistically grounded, generalizable, and accessible tools for biological discovery.

\begin{figure}[!t]
    \centering
    \includegraphics[width=1\linewidth]{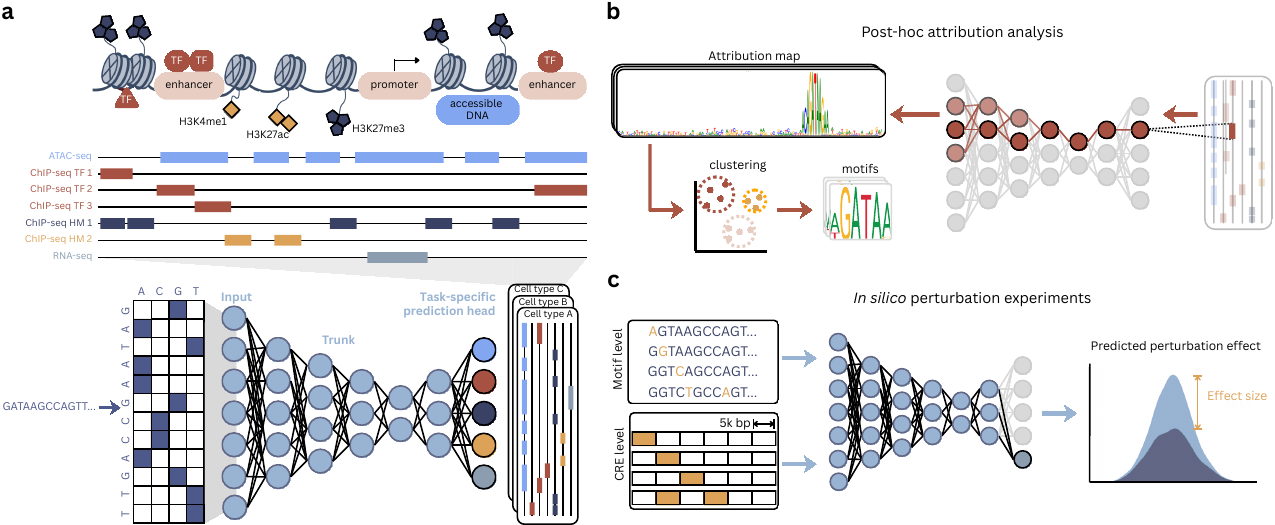}
    \caption{Standard seq2func workflow. \textbf{a}, (Top) Schematic of gene regulation and functional genomics tracks. (Bottom) A seq2func model takes one-hot–encoded DNA sequence as input, processes it through a feedforward trunk of shared layers, and produces predictions via task-specific heads, which may be single-task or multi-task. \textbf{b}, Schematic of post hoc attribution analysis, in which gradients or related signals are propagated from a selected output track back to the input sequence to generate an attribution map. Collections of attribution maps can be clustered to identify sequence motifs as contribution weight matrices. \textbf{c}, Schematic of an \textit{in silico} experimentation platform. DNA sequences are perturbed at multiple scales---ranging from single-nucleotide variants to motif-level or \textit{cis}-regulatory–element perturbations---and evaluated by the model to predict quantitative effect sizes. Perturbations introduced within a locus assess necessity, while elements placed in diverse sequence contexts and averaged assess sufficiency.}
    \label{fig:seq2func_workflow}
\end{figure}

\section*{Modeling Approaches for Regulatory Genomics} 

Genomic sequence modeling broadly spans supervised seq2func models, unsupervised or self-supervised sequence models, and conditional generative models. Seq2func models are trained directly on experimentally measured functional readouts, including TF binding, chromatin accessibility, and gene expression, from DNA sequence \cite{eraslan2019deep} (Fig.~\ref{fig:seq2func_workflow}a). Unsupervised models are trained without functional labels and instead learn representations from sequence statistics alone. Functional information is introduced only after pretraining, either by fine-tuning on labeled data, training downstream predictors on learned embeddings, or guiding sequence generation with external scoring functions. This class is currently dominated by genomic language models (gLMs) \cite{benegas2025genomic}. Conditional generative models learn sequence distributions explicitly conditioned on functional or contextual variables, integrating supervision directly into the generative objective rather than appending it after pretraining. This formulation appears in several diffusion-based frameworks and in a subset of conditionally trained gLMs \cite{dasilva2025designing,avdeyev2023dirichlet,stark2024dirichlet,sarkar2024designing,lal2024designing}.

Despite growing interest in generative approaches, supervised seq2func models remain the primary workhorse in regulatory genomics. This contrasts with protein sequence modeling, where unsupervised training has been effective \cite{lin2023evolutionary,frazer2021disease}. Protein sequences evolve under strong positional and covariation constraints, so sequence variation across evolutionary distances strongly reflects sequence–structure relationships \cite{morcos2011direct}. In regulatory DNA, evolutionary constraints act primarily on regulatory output rather than on specific nucleotide identities, leading to weak and diffuse sequence-level constraints \cite{dermitzakis2002evolution}. Many distinct sequences can therefore encode similar regulatory activity, while the sequence features that determine function remain sparse relative to background \cite{kim2023deciphering}. Regulatory activity is also intrinsically context-dependent, so the same sequence can drive different outcomes across cell types. Consequently, reconstruction-based objectives are dominated by background sequence variation, overwhelming the sparse sequence features that determine regulatory function. Consistent with this, gLMs pretrained on whole genomes often learn less informative representations in human regulatory regions than seq2func models \cite{tang2025evaluating,patel2024dart}.

Several strategies have been proposed to steer unsupervised learning toward regulatory features, including restricting training to annotated regulatory regions or incorporating evolutionary conservation \cite{lal2024designing,ye2025predicting,benegas2025dna}. Their effect on learning \textit{cis}-regulatory logic, however, has not been rigorously established. Discrete-diffusion approaches offer a promising alternative by modeling mutational transitions that evolve random sequences toward functional ones \cite{sarkar2024designing,chandra2025unification}, avoiding reconstruction-centered training objectives, though current implementations remain limited in scale compared with gLMs.

By contrast, seq2func models leverage large-scale functional genomics datasets spanning multiple assays, cell types, and conditions, enabled in part by consortia such as ENCODE \cite{encode2012integrated}. These datasets provide high-throughput functional signals, allowing seq2func models to learn sequence features that predict quantitative regulatory activity across cellular contexts.

\section*{Architectural Biases for Multiscale Modeling} 

Seq2func models are best viewed as function approximators that learn a mapping from DNA sequence to regulatory activity measured by a functional assay \cite{cybenko1989approximation}. In principle, any sufficiently expressive architecture could approximate this mapping given unlimited data. In practice, regulatory genomics operates in an extreme finite-data regime: the space of possible sequences is vast, while experiments sample only a tiny fraction of it. Generalization therefore depends critically on inductive bias, which constrains the class of functions a model can learn from limited observations. When successful, a trained seq2func model can be treated as a \emph{virtual assay}, enabling \textit{in silico} measurements of novel sequences under matched experimental conditions.

Here we focus on architectural bias, which governs how information is propagated across positions and layers. A useful concept is the receptive field, which represents the span of input positions that influence a prediction. Convolutional layers aggregate information locally, with longer-range interactions emerging only as depth expands the receptive field. Dilated convolutions and pooling layers accelerate this expansion while preserving locality consistent with motif syntax. Although individual convolutional layers behave additively across positions, deeper stacks can capture higher-order dependencies, including flanking-sequence effects and motif--motif interactions \cite{koo2021global,avsec2021base,de2022deepstarr,horton2023short}.

Other architectures support more direct long-range integration. Recurrent models, such as bidirectional LSTMs \cite{quang2016danq}, aggregate context sequentially but are constrained by optimization stability and computational cost. These limitations motivated increased interest in transformers and state-space models. Transformers use multi-head self-attention to capture pairwise interactions across the full sequence, and by stacking multiple layers, compose these interactions into complex combinatorial dependencies \cite{vaswani2017attention}, but this expressivity typically increases data requirements. State-space architectures, such as Hyena \cite{poli2023hyena} and Mamba \cite{gu2023mamba}, propagate information through structured recurrent or gating dynamics with linear-time scaling. Although fully connected layers can, in principle, capture interactions across the full sequence, their lack of structural bias makes them data inefficient for regulatory sequences. Accordingly, dense layers are typically used only after earlier layers extract localized features.

Because the \textit{cis}-regulatory code spans multiple length scales, seq2func architectures are composed of stacked layers to capture its hierarchical organization. Early convolutional layers aim to capture motif-level patterns \cite{koo2019representation}, while later attention or state-space layers integrate these features to model distal regulatory interactions. Additional inductive bias can be imposed through architectural constraints. Squeeze–excitation modules provide context-dependent channel reweighting \cite{hu2018squeeze}, residual connections support contextual refinement of early features \cite{he2016deep}, and symmetry constraints such as reverse-complement weight sharing reduce redundancy \cite{shrikumar2017reverse,mallet2021reverse}.

Architectural comparisons rarely yield decisive winners. Performance differences often reflect cumulative implementation choices rather than inherent architectural superiority. Broad searches over architectures, training protocols, and hyperparameters routinely produce multiple models with similar accuracy, with gains driven by many incremental refinements rather than single innovations. Hence, uneven tuning effort can confound model comparisons. In some settings, well-tuned convolutional networks match the performance of more expressive transformer-based models \cite{rafi2025community,yang2024convolutions,liu2022convnet}.

Conceptually, architecture and training data jointly define the loss landscape explored during optimization. While architectural changes may not alter representational capacity in principle, they can substantially reshape the geometry of this landscape, making certain solutions easier to reach under current initialization and optimization strategies \cite{glorot2010understanding,he2015delving,kingma2014adam}. Evaluation should therefore go beyond peak predictive performance and emphasize data efficiency as a probe of inductive bias, together with sensitivity to hyperparameter choices as a measure of robustness.

\section*{Data and Task Design Shape Seq2func Models} 

Seq2func prediction tasks rely primarily on two classes of functional genomics data: genome-wide profiling assays and massively parallel reporter assays (MPRAs)\cite{kinney2019massively}. Profiling assays measure regulatory activity in endogenous context, with models trained to predict signal intensity within a local window, learning sequence–activity relationships from genetic variation across the genome. MPRAs instead measure the activity of large collections of natural or synthetic sequences under matched experimental conditions, either episomally or following genomic integration via lentiviral delivery \cite{inoue2017systematic}.

Early seq2func models framed regulatory activity as binary classification based on peak calls \cite{zhou2015predicting,kelley2016basset}. More recent work has shifted toward quantitative profile prediction, in which models regress full signal tracks across genomic windows \cite{avsec2021base,kelley2018sequential}. Quantitative supervision captures richer aspects of \textit{cis}-regulatory organization than binary labels, including motif strength, footprint structure, and local chromatin context, and substantially improves SNV-effect prediction \cite{toneyan2022evaluating,pampari2025chrombpnet,avsec2025alphagenome}. Consistent with this, profile-trained models learn motif representations whose quantitative preferences align with biophysical measurements \cite{horton2023short}.

\begin{figure}
    \centering
    \includegraphics[width=0.5\linewidth]{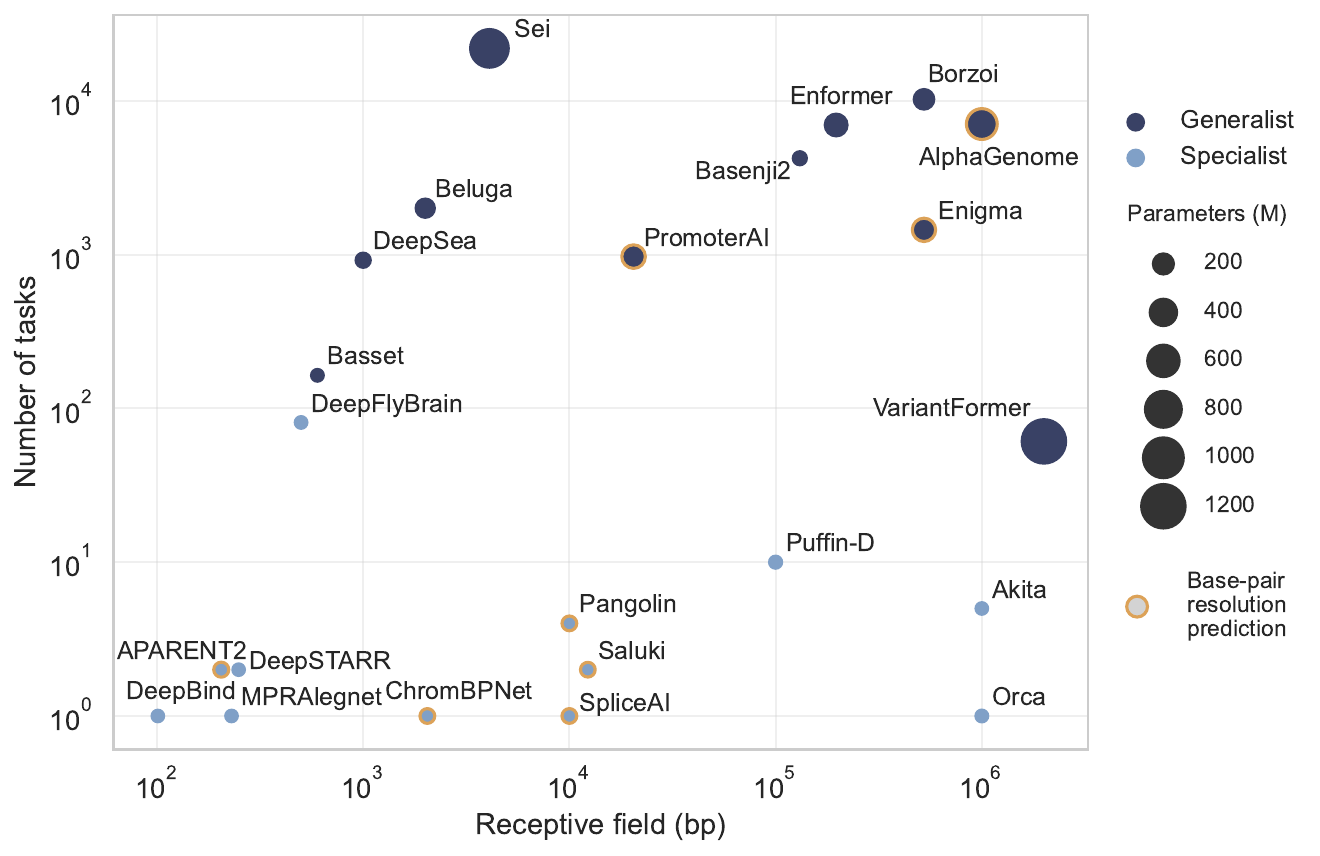}
    \caption{Landscape of seq2func models by genomic receptive field and task breadth. Shown is the number of prediction tasks versus the input receptive field for representative seq2func models. Colors indicate model type: generalist models correspond to multitask architectures (navy), while specialist models correspond to single-task architectures (light blue). Marker size is proportional to the reported parameter count. A gold marker edge indicates models that produce base-pair–aligned predictions. Select models shown include: DeepSea \cite{zhou2015predicting}, Basset \cite{kelley2016basset}, Beluga \cite{zhou2018deep}, Sei \cite{chen2022sequence}, Basenji2 \cite{kelley2018sequential}, PromoterAI \cite{jaganathan2025predicting}, Enformer \cite{avsec2021effective}, VariantFormer \cite{ghosal2025variantformer}, Enigma \cite{jung2025enigma}, Borzoi \cite{linder2025predicting}, AlphaGenome \cite{avsec2025alphagenome}, DeepBind \cite{alipanahi2015predicting}, APARENT2 \cite{linder2022deciphering}, MPRA-LEGNet \cite{agarwal2025massively}, DeepSTARR \cite{de2022deepstarr}, ChromBPNet \cite{pampari2025chrombpnet}, SpliceAI \cite{jaganathan2019predicting}, Saluki \cite{agarwal2022genetic}, Pangolin \cite{zeng2022predicting}, Puffin-D \cite{dudnyk2024sequence}, Akita \cite{fudenberg2020predicting}, and Orca \cite{zhou2022sequence}.}
    \label{fig:seq2func_models}
\end{figure}

Seq2func models may be trained as specialists for a single assay or as generalists across assays or cell types in a multitask setting. Single-task models like chromBPNet \cite{pampari2025chrombpnet} are optimized for one experimental condition and often perform well in-distribution, but degrade under distribution shift. Multitask models share supervision across experiments, allowing them to learn regulatory features that recur across conditions while benefiting from larger effective training sets \cite{linder2025predicting,avsec2025alphagenome,boshar2025foundational}. Recent trends emphasize scaling both model capacity and training data diversity, supporting foundation-style models that operate over longer sequences and generalize across assays, cell types, and species (Fig.~\ref{fig:seq2func_models}).

In practice, multitask learning introduces characteristic biases that are often masked by aggregate performance metrics. Because many regulatory programs are shared across cell types, optimization can favor shared features that are broadly predictive across tasks, such as housekeeping regulatory programs. These shared signals can dominate the loss, leading models to underrepresent features that drive differential regulation across contexts, particularly among closely related cell types. As a result, models may achieve strong overall accuracy while struggling to capture the regulatory logic underlying context-specific activity \cite{kathail2024current}. This limitation motivates evaluation stratified by differentially regulated elements. Although targeted fine-tuning, data upsampling, and focal-style losses can partially recover context-specific signals, a general solution remains an open challenge.

\section*{Interpreting \textit{Cis}-Regulatory Insights from Seq2Func Models} 

Despite strong predictive performance, seq2func models provide limited direct insight into the regulatory mechanisms underlying their predictions. Interpreting the sequence features a model uses, their interactions, and their context dependence relies on post hoc attribution methods, virtual perturbation experiments, and analyses of internal representations.

\paragraph{Local attributions.}
Post hoc attribution methods estimate the contribution of individual nucleotides to a model's prediction for a given sequence (Fig.~\ref{fig:seq2func_workflow}b). Common approaches include gradient-based methods \cite{simonyan2013deep, sundararajan2017axiomatic}, additive contribution methods \cite{shrikumar2017learning, lundberg2017unified}, and perturbation-based schemes such as \textit{in silico} mutagenesis (ISM) \cite{zhou2015predicting}. Attribution methods differ in whether they probe local sensitivity or estimate additive contributions, and in how the local sequence neighborhood is defined \cite{han2022explanation}. These methods require a scalar output. For profile-predicting models, the choice of how a profile is reduced to a scalar can strongly affect the resulting attributions. PISA avoids this ambiguity by providing attribution maps at individual positions of the predicted profile \cite{mcanany2025pisa}.

The reliability of attribution methods depends strongly on the learned sequence–activity function. Because seq2func models are optimized for predictive accuracy rather than local smoothness, locally rough sequence–activity functions arising from noise or benign overfitting can render gradient-based attributions unstable, even when held-out performance is strong \cite{alvarez2018robustness,majdandzic2022selecting}. The discrete nature of DNA further complicates interpretation, as gradients are computed off the categorical sequence manifold, introducing spurious components that appear as noise in attribution maps \cite{majdandzic2023correcting}. Statistical gradient correction and gauge-fixing approaches can mitigate these effects \cite{majdandzic2023correcting,posfai2025gauge}.

\paragraph{From additivity to interactions.} 
Attribution maps are often summarized as sequence logos, providing a first-order view of motifs and local syntax. However, additive attributions cannot capture dependencies arising from interactions among bases or motifs, including cooperativity, spacing effects, and context dependence. Pairwise interactions can be probed using perturbation-based strategies such as second-order ISM \cite{koo2018residualbind}, hybrid perturbation–attribution methods such as DFIM \cite{greenside2018discovering}, or second-order gradient methods like Hessians \cite{janizek2021explaining}. Interpretation requires care, as apparent interactions may reflect non-additivity or aggregated additive effects. This ambiguity can be resolved using gauge-fixing approaches that separate additive and interaction components without altering the underlying function \cite{posfai2025gauge}, or through statistical testing frameworks such as DIAMOND \cite{chen2025error}. 

\paragraph{Global explanations.}
Attribution maps are inherently sequence-specific. Identifying patterns that recur across many sequences is therefore critical for uncovering global mechanistic structure (Fig.~\ref{fig:seq2func_workflow}b). TF-MoDISco clusters recurrent high-attribution subsequences into motif-like patterns \cite{shrikumar2018technical}. SEAM instead clusters full attribution maps derived from dense mutagenesis libraries of a reference sequence, revealing the repertoire of regulatory mechanisms accessible through a few, key mutations \cite{seitz2025uncovering}. This analysis uncovers \textit{de novo} motif emergence, context dependence, and cooperative or competitive interactions. More broadly, classical sequence analysis tools can be adapted to attribution maps to extract mechanistic structure.

\paragraph{Counterfactual probing through virtual perturbations.}
\textit{In silico} perturbation analysis treats seq2func models as surrogates for functional assays, enabling controlled tests of cause-effect relationships between sequence features and predicted regulatory activity \cite{koo2020deeptfbs}. Removing a feature tests necessity, inserting a feature into randomized sequences tests sufficiency, and relocating features probes positional and spacing effects \cite{koo2021global, avsec2021base,de2022deepstarr,toneyan2022evaluating,gunsalus2023silico,toneyan2024interpreting} (Fig.~\ref{fig:seq2func_workflow}c). Combinatorial logic is probed by comparing single-feature versus multi-feature perturbations to test for non-additivity. To avoid confounding from spurious sequence patterns, perturbations are applied across randomized backgrounds and averaged. This marginalization strategy, formalized in global importance analysis \cite{koo2021global}, isolates the causal effect of the perturbed pattern through the lens of the model. Recent toolkits streamline the design and execution of these perturbation campaigns \cite{klie2023eugene,schreiber2025tangermeme}.

Dense perturbation schemes further enable surrogate modeling approaches that fit structured, interpretable functions to approximate seq2func predictions. SQUID adapts local surrogate modeling to genomics by fitting quantitative models whose parameters correspond directly to \textit{cis}-regulatory features and interactions \cite{seitz2024interpreting}. In the additive limit, surrogate parameters resemble attribution scores, but the framework naturally extends to pairwise and biophysical models. However, computational costs restrict these analyses to focused loci.

\paragraph{Inherently interpretable representations.}
Architectural constraints can bias learned parameters toward biologically meaningful structure. Unconstrained first-layer convolutional filters do not reliably learn motifs; stable motif representations emerge only when strong architectural biases are introduced, such as large pooling or highly nonlinear activations that amplify regulatory signals while suppressing background \cite{koo2019representation,koo2021improving}. In deeper layers, attention mechanisms can reveal putative motif–motif interactions \cite{ullah2021self}, offering a higher-level view of regulatory organization, although attention maps for individual sequences are often noisy. Aggregation frameworks such as GLIFAC summarize attention-weighted motif co-occurrences across many sequences, providing global motif-motif interactions \cite{ghotra2021uncovering}.

\paragraph{Clarifying complex representations.}
Sparse autoencoders (SAEs) decompose uninterpretable dense embeddings into sparse latent features intended to capture interpretable concepts \cite{elhage2022toy}. SAEs have recently been applied to protein language models \cite{simon2025interplm,adams2025mechanistic} and genomic-based models \cite{brixi2025genome,maiwald2025decode,korsakova2025learning}. Although SAEs can be trained reliably at scale, individual latents often capture fragmented or merged signals (i.e., feature splitting and absorption) rather than well-defined biological concepts \cite{leask2025sparse,peng2025use}. Consequently, interpretation typically relies on indirect evidence, such as motif enrichment, overlap with genomic annotations, or correlations with functional tracks. Virtual perturbation experiments can provide a more direct test of whether SAE latents correspond to specific regulatory mechanisms.

\paragraph{Outlook.} 
No single interpretability method provides a complete explanation of a seq2func model. Attribution methods support hypothesis generation, while virtual perturbation experiments enable causal probing of learned mechanisms. However, perturbation-based analyses depend on model reliability under distribution shift, which is not guaranteed. Consequently, misleading interpretations may arise not from the interpretability method itself, but from poor model generalization. Meaningful insights therefore requires employing complementary interpretability strategies.

At present, extracting insight remains labor-intensive and human-driven, limiting scalability. Progress will require automated systems that can reason over large collections of attribution and perturbation results while integrating prior biological knowledge. Recent advances in large language models suggest two complementary directions. One is agentic AI systems \cite{gao2024empowering} that orchestrate existing interpretability tools to identify recurring regulatory patterns, synthesize results, generate mechanistic hypotheses, and propose targeted virtual perturbations. The other is LLM-guided global surrogate modeling, in which interpretable functions are discovered and refined through closed-loop program-search procedures, as exemplified by AlphaEvolve \cite{novikov2025alphaevolve} and LLMGEN \cite{rajesh2025llmgen}. Together, these approaches point toward a shift from manual interpretation to more automated discovery of \textit{cis}-regulatory rules.

\begin{table}[!b]
\centering
\caption{Distribution shifts in seq2func model deployment and their implications for evaluation.
Shown are common deployment scenarios for seq2func models, the corresponding forms of distribution shift (covariate, label, and concept), what changes in each setting, and representative applications that motivate perturbational evaluations to assess generalization.}
\label{tab:distribution_shifts}
\footnotesize
\renewcommand{\arraystretch}{1.25}
\setlength{\emergencystretch}{3em}
\begin{tabular}{p{2.6cm}p{1.2cm}p{3.0cm}p{3.5cm}}
\toprule
\textbf{Scenario} & \textbf{Shift Type} & \textbf{What Changes} & \textbf{Application} \\
\midrule
Genetic variants & Covariate & SNVs and small indels & Personalized genomes, rare disease variant interpretation \\
\hline
Motif rearrangement & Covariate & Regulatory grammar and motif organization & \textit{cis}-regulatory grammar inference, rational enhancer design \\
\hline
Evolutionary \newline rearrangements & Covariate & Large-scale structural variants: translocations, inversions, rearrangements & Cancer structural variation interpretation, cross-species regulatory transfer \\
\hline
Synthetic sequences & Covariate & \textit{De novo} sequences outside natural genomic sequence space & De novo regulatory sequence design, synthetic promoter/enhancer engineering \\
\specialrule{1.1785pt}{0pt}{0pt}
Cross-experiment & Label & Same underlying biology, different measurement processes and technical biases &  Assay transfer and cross-platform functional genomics integration \\
\specialrule{1.1785pt}{0pt}{0pt}
Cross-cell-state \newline and cell-type & Concept & Changes in the sequence-function mapping due to cellular context or signaling-driven state changes & Context-specific regulatory modeling in development, disease, and drug response \\
\bottomrule
\end{tabular}
\end{table}

\section*{Evaluating Generalization in Seq2Func Models} 

The democratization of deep learning has lowered the barrier to building seq2func models, accelerating exploration of architectures and training strategies. However, model development has outpaced progress in rigorous evaluation. This gap is particularly consequential because seq2func models are trained on experimentally measured regulatory activity but are primarily deployed to predict the behavior of unassayed sequences. Evaluation must therefore probe generalization beyond the training distribution.

A useful lens for formalizing this challenge is distribution shift, which arises when data encountered at deployment differ from those observed during training \cite{moreno2012unifying}. In regulatory genomics, three forms are especially relevant: covariate shift, label shift, and concept shift. Each arises naturally in downstream applications and poses distinct challenges for evaluation.

\textit{Covariate shift} occurs when the input sequence distribution changes while the underlying sequence–function mapping remains fixed. This is the most common shift encountered in seq2func deployment. SNV effect prediction typically involves small, local perturbations around the reference genome, whereas analyses of variant combinations, motif rearrangements, regulatory grammar edits, or structural variants require extrapolation to increasingly distant regions of sequence space. \textit{De novo} regulatory sequence design represents the extreme, probing sequences far outside the natural genomic distribution. Together, these settings span a continuum of covariate shift in which test sequences progressively diverge from those seen during training (Table~\ref{tab:distribution_shifts}).

Despite this, most evaluation protocols weakly probe covariate shift. Common benchmarks rely on random splits or chromosome holdouts from a single experiment, which preserve substantial local and compositional sequence similarity. Such strategies do not prevent information leakage: compositionally similar sequences can reside on different chromosomes, allowing homologous signals to be shared between training and test sets \cite{rafi2025detecting}. As a result, reported performance can overestimate generalization to genuinely novel sequences.

\textit{Label shift} arises when the input distribution is similar, but different assays measure the same underlying biological activity and produce systematically different observed measurements. This is common in cross-assay or zero-shot evaluations, where models trained on one experiment are tested against another \cite{critical2024cagi,karollus2023current,huang2023personal,sasse2023benchmarking}. Examples include predicting MPRA perturbation effects \cite{melnikov2012systematic}, CRISPR-based screens \cite{qi2013repurposing}, or statistical associations such as chromatin accessibility or expression QTLs \cite{degner2012dnase}. While informative, these comparisons are difficult to interpret because assays differ in noise structure, bias, and dynamic range, even when probing the same underlying biology.

Some assay-specific distortions, such as Tn5 insertion bias in ATAC-seq \cite{buenrostro2013transposition}, can be partially corrected \cite{martins2018universal,pampari2025chrombpnet}, but many sources of label shift remain poorly characterized. More generally, models that explicitly consider the measurement process aim to disentangle latent regulatory activity from assay-specific readouts \cite{otwinowski2018global,tareen2022mave}. Apparent cross-assay performance differences may therefore reflect properties of the measurement process rather than differences in the underlying biology \cite{sailer2017detecting}.

\textit{Concept shift} represents a more severe regime in which the sequence–function mapping itself changes. This occurs when models trained in one cell type, cellular state, or species are applied to another, where TF concentrations, chromatin state, and regulatory programs differ. The severity of concept shift depends on biological divergence. 

Distinguishing failures due to distributional shift is therefore essential for interpreting benchmark results. Meaningful progress requires sustained community coordination around shared benchmarks that reflect realistic deployment settings, employ well-defined metrics tied to biologically interpretable outcomes, and specify clear criteria for success. In the absence of such shared objectives, evaluation remains fragmented, with models assessed on disparate datasets in ways that make progress difficult to measure.

Emerging efforts such as GAME \cite{luthra2025game} represent important steps toward standardized evaluation. Looking forward, benchmark design should emphasize perturbational data collected in controlled biological systems and span progressively larger distributional shifts, enabling performance to be explicitly calibrated to downstream use cases. Other fields have advanced through long-running, community-driven benchmarks with stable objectives, such as ImageNet in computer vision \cite{russakovsky2015imagenet} and CASP in protein structure prediction \cite{yuan2026casp16}. Despite several recent competitions in regulatory genomics \cite{encode_dream2017tf,critical2024cagi,rafi2025community}, the field still lacks a unifying benchmark of comparable scope.

\begin{figure}[!b]
    \centering
    \includegraphics[width=1\linewidth]{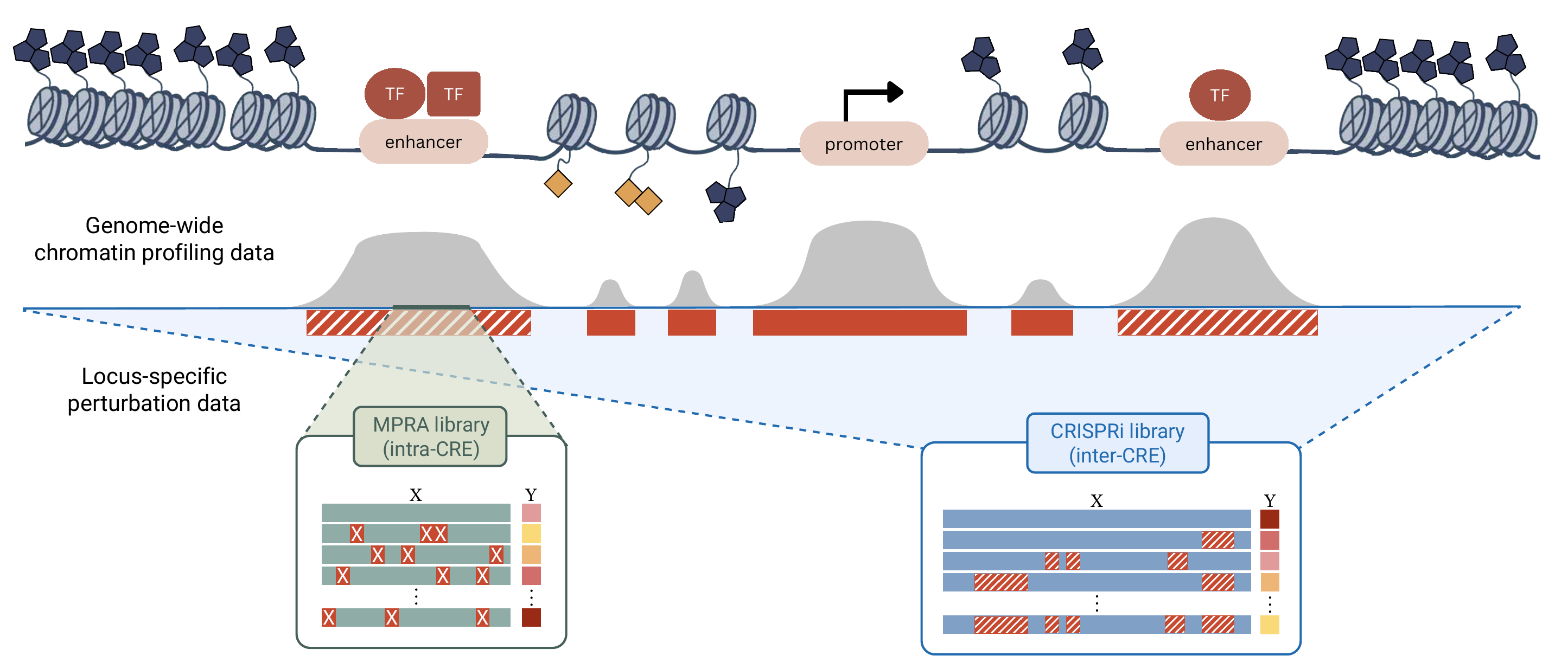}
    \caption{Perturbation libraries to probe the \textit{cis}-regulatory code. Massively parallel reporter assays (MPRAs) introduce dense sequence variation within \textit{cis}-regulatory elements (CREs) to measure how local nucleotide changes affect regulatory activity. In contrast, CRISPR interference (CRISPRi) libraries target entire CREs, individually or in combination, within their native chromatin context, enabling tests of CRE necessity, CRE–CRE interactions, and long-range or context-dependent regulatory effects.}
    \label{fig:perturbation_library}
\end{figure}

\section*{Generalization Gap in Seq2Func Models} 

Despite substantial progress, robust generalization remains a central challenge for seq2func models. This limitation is driven less by model architecture or capacity than by the structure of available training data. Most seq2func models, including large multitask systems, are trained on genome-wide functional genomics assays aligned to a reference genome \cite{avsec2021effective,linder2025predicting,avsec2025alphagenome}. This training regime samples only a narrow and biased region of regulatory sequence space, even at genome scale. Consequently, strong performance on held-out genomic regions does not guarantee reliable predictions under novel perturbations, engineered sequences, or alternative regulatory configurations \cite{sasse2023benchmarking,huang2023personal,karollus2023current}. This reflects a mismatch between the breadth of \textit{cis}-regulatory mechanisms we seek to model and the limited coverage present in current training data.

One response has been to increase data density in regions of sequence space relevant to downstream applications. Motivated by weak performance on personal variant prediction, several studies fine-tune pretrained genome-wide models on paired personal genomic and transcriptomic data \cite{rastogi2024fine,drusinsky2024deep,spiro2025scalable}. While this improves predictions for represented genes and individuals, it generalizes poorly to unseen genes or populations. This limitation reflects the structure of human genetic variation: ancestry, linkage, and purifying selection constrain variants to narrow, highly correlated neighborhoods around the reference genome. Consequently, personal variation provides limited leverage for inferring causal regulatory rules that generalize across broader sequence space. Cross-species sequence diversity from resources such as Zoonomia \cite{zoonomia2020comparative} and the Vertebrate Genomes Project \cite{rhie2021towards} offers additional variation, though its impact on regulatory generalization remains uncertain.

More fundamentally, genome-wide profiling assays only sample from naturally occurring genetic variation, spanning an infinitesimal fraction of regulatory sequence space. Models trained on this sparse and biased sampling can achieve strong in-distribution performance while lacking the resolution needed for extrapolation. In principle, robust generalization would emerge if models learned causal or biophysical rules of regulation. In practice, models trained primarily on observational data tend to capture coarse statistical associations, yielding a ``blurry vision'' of the regulatory genome rather than the fine-grained mechanistic understanding required to predict novel variation.

One route to improved generalization is to encode causal or biophysical constraints directly into model architectures, but this approach is limited by incomplete mechanistic knowledge. A more promising alternative is to train on data that explicitly encode causal information. Perturbation assays, such as MPRAs and CRISPR-based screens, achieve this by measuring regulatory responses to controlled sequence edits (Fig.~\ref{fig:perturbation_library}). Extracting mechanistic rules requires dense perturbation sampling that spans diverse mechanisms and a wide dynamic range of activity \cite{kinney2019massively}. Despite advances in oligonucleotide synthesis and pooled screening, such dense perturbations remain feasible only at individual loci or small locus sets.

Hence, perturbation assays provide high-resolution views of regulatory mechanisms but limited genomic coverage, whereas genome-wide profiling assays provide broad coverage but low mechanistic resolution. Bridging this resolution–coverage tradeoff in a way that supports both genome-wide generalization and fine-grained mechanistic learning remains a central obstacle for seq2func modeling.

\begin{figure}[!b]
    \centering
    \includegraphics[width=.9\linewidth]{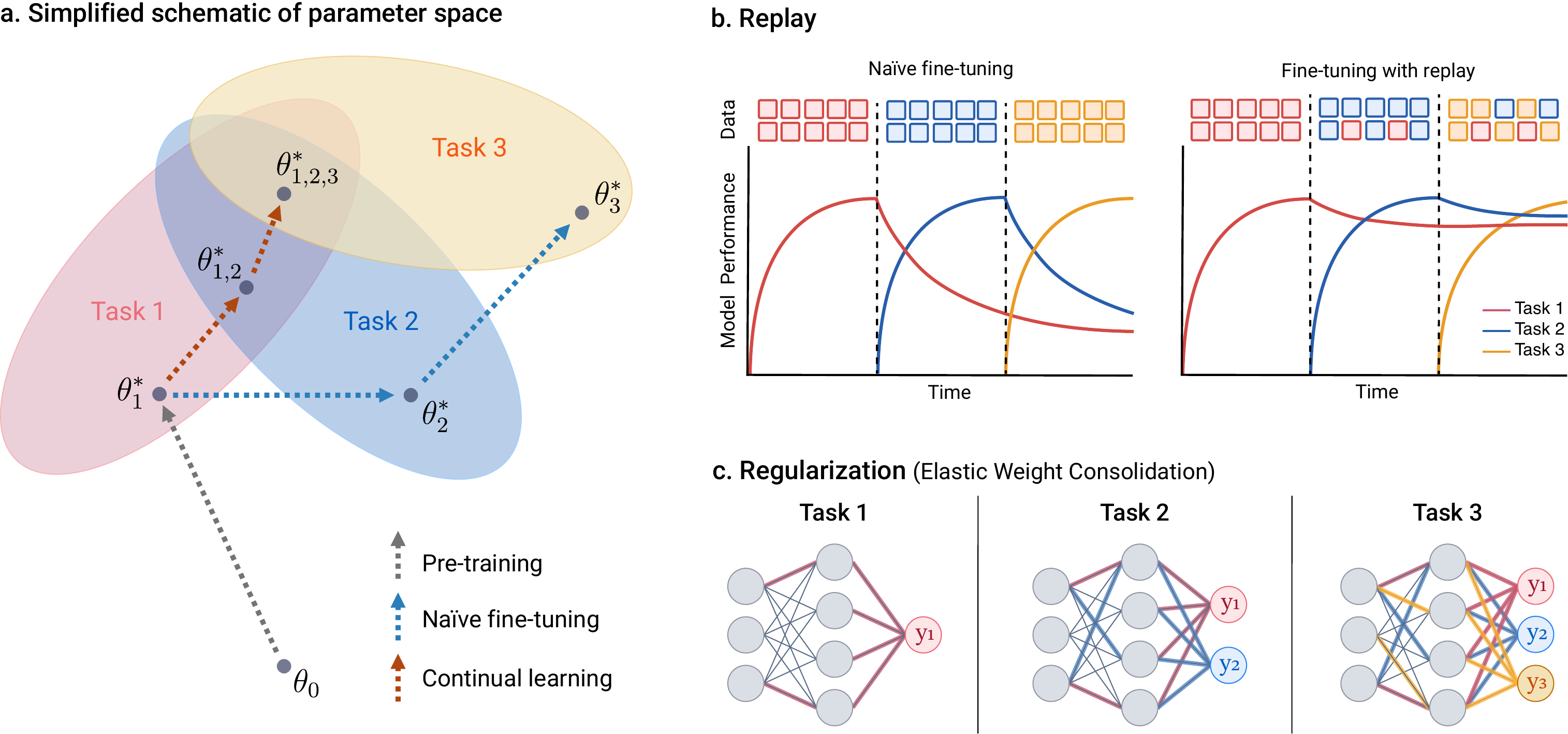}
    \caption{Overview of continual learning strategies. \textbf{a}, A conceptual view of parameter space shows how naïve fine-tuning pushes parameters toward the optimal shaded region of each new task, often degrading performance on earlier tasks, whereas continual learning methods encourage movement toward shared optima that preserve prior capabilities. \textbf{b}, Replay-based methods maintain performance over time by mixing past examples with new data during fine-tuning. \textbf{c}, Regularization-based methods, such as Elastic Weight Consolidation, reduce forgetting by constraining updates to parameters deemed important for earlier tasks.}
    \label{fig:continual_learning}
\end{figure}

\section*{Continual Learning Across Genomic Assays} 

A natural strategy for integrating genome-wide profiling assays with locus-specific perturbation data is joint multitask training or fine-tuning. In practice, joint training is limited by extreme data imbalance: dense perturbation experiments generate thousands of measurements at a small number of loci, whereas most of the genome is observed only once. Even with loss reweighting or sampling heuristics, optimization is dominated by a few overrepresented regions, biasing models toward locus-specific effects rather than transferable regulatory principles. Naive fine-tuning introduces a complementary failure mode. Fine-tuning a pretrained model on perturbation data can improve predictions at the targeted locus, but updating shared parameters risks catastrophic forgetting \cite{french1999catastrophic}, degrading performance elsewhere in the genome. Repeating this process across loci yields a collection of locus-specialized models rather than a single model that accumulates generalizable mechanistic insight.

These limitations motivate a \emph{continual learning} formulation, in which models are updated incrementally as new data arrive while preserving previously acquired capabilities \cite{kirkpatrick2017overcoming,wickramasinghe2023continual,wang2024comprehensive} (Fig.~\ref{fig:continual_learning}a). Continual learning methods differ in how they mitigate catastrophic forgetting. Replay-based approaches interleave stored or generated examples from prior tasks to approximate the original training distribution \cite{rolnick2019experience} (Fig.~\ref{fig:continual_learning}b). Regularization-based methods, such as elastic weight consolidation, penalize updates to parameters estimated to be important for earlier tasks \cite{kirkpatrick2017overcoming} (Fig.~\ref{fig:continual_learning}c). Architecture-based strategies reduce interference by allocating task-specific parameters or modules. More recently, nested learning frames adaptation as a hierarchy of coupled optimization processes \cite{behrouz2025nested}, though their practical advantages over standard continual learning methods remain unclear.

In overparameterized models, many parameter configurations achieve similar performance, often residing in relatively flat regions of the loss landscape \cite{li2018visualizing}. Continual learning seeks to exploit weakly constrained directions in parameter space (\textit{sloppy modes}) that lie near flat minima, allowing parameters to shift to accommodate new tasks while preserving performance on previously learned ones. Which strategies, or combinations thereof, best support integration of perturbation data into genome-wide models remains an open question.

Applying continual learning to regulatory genomics introduces challenges not encountered in more homogeneous domains such as vision or language. Perturbation datasets vary widely in scale, resolution, modality, and biological scope, spanning single-nucleotide edits, motif-level perturbations, and long-range \textit{cis}-regulatory element effects. Effective continual learning must therefore accommodate not only sequential data acquisition, but also shifts in biological scale, assay modality, measurement bias, and technical noise.

Perturbation experiments do not define isolated tasks at independent loci. Instead, they reveal \textit{cis}-regulatory mechanisms that recur across the genome, making locus-by-locus refinement conducive to knowledge transfer. Insight gained at one locus can improve predictions at others with similar regulatory architectures (\emph{forward transfer}), while refinement on new loci can sharpen previously learned mechanisms (\emph{backward transfer}) \cite{lopez2017gradient,kemker2018measuring}. The central challenge is to convert local mechanistic insight from targeted perturbations into coherent, genome-wide regulatory knowledge. In this view, perturbation datasets serve as mechanistic constraints that progressively refine a model’s understanding of \textit{cis}-regulatory rules.

\section*{Designing Informative Perturbations for Generalization} 

Because perturbation assays are costly and sequence space is vast, a central challenge is designing perturbations that most effectively improve generalization. Ideally, perturbation datasets would sample sequence space in ways that probe diverse regulatory mechanisms while producing a broad and well-balanced dynamic range of activities. In practice, perturbation design remains largely heuristic, driven by experimental convenience or interpretability rather than principled estimates of information gain.

Saturation mutagenesis introduces single-nucleotide changes and is widely used for its tractability \cite{patwardhan2009high}, but primarily probes additive effects and offers limited insight into motif syntax or higher-order interactions. Partial random mutagenesis perturbs multiple positions near a parent sequence, enabling sampling of more complex interactions, but yields datasets that are difficult to interpret without explicit modeling \cite{kinney2010using}. Motif-centric designs manipulate known TF motifs to test combinatorial logic, but are inherently biased toward previously characterized mechanisms \cite{sahu2022sequence,de2022deepstarr,friedman2025active}. At the opposite extreme, fully random libraries sample sequence space broadly \cite{de2024hold}, but their utility is highly context dependent: random sequences can drive expression in yeast \cite{de2020deciphering,vaishnav2022evolution}, yet are largely inactive in mammalian systems \cite{luthra2024regulatorydefault,camellato2024syntheticdefault}, producing highly imbalanced readouts.

These limitations motivate intermediate strategies that balance sequence diversity with biological plausibility. Structured, evolution-inspired perturbations introduce diversity across multiple scales while preserving realistic sequence organization. Genome-Shuffle-seq \cite{pinglay2025multiplex} achieves this through deletions, inversions, translocations, and mutations. Related evolution-inspired data augmentations via EvoAug have proven effective \textit{in silico} \cite{lee2023evoaug,duncan2024improving,yu2024evoaug}, suggesting that experimentally measured perturbations of this form could provide especially informative training signals.

\emph{Active learning} offers a complementary, principled framework by prioritizing perturbations expected to yield maximal information gain \cite{gal2017deepbal}. In regulatory genomics, this problem is naturally posed as batch-mode active learning \cite{kirsch2019batchbald,ash2020badge}, in which sets of sequences are selected jointly rather than individually. Common acquisition strategies emphasize uncertainty, model disagreement, or sequence diversity, but these proxies are often weakly coupled to information gain \cite{mackay1992information,settles2009active}. For example, models may assign high uncertainty to compositionally extreme and diverse sequences, such as homopolymeric sequences, yet provide little insight into regulatory mechanisms.

Unlike classical active learning, regulatory genomics lacks a fixed pool of unlabeled data \cite{settles2009active}. Experimental design therefore involves two coupled steps: generating a candidate reservoir of sequences, and selecting an informative batch for measurement via an acquisition function. Reservoirs may be constructed through perturbation schemes, sequence-design methods \cite{taskiran2024cell,gosai2024machine,schreiber2025programmatic}, or generative models \cite{avdeyev2023dirichlet,sarkar2024designing,king2025generative,boshar2025foundational}. Failures can arise from either component: a reservoir that fails to expose informative mechanisms, or an acquisition rule that selects sequences that are uncertain but mechanistically uninformative. Design choices must also balance informativeness against cost: fully synthesized libraries are expensive, random libraries are inefficient but cheap to synthesize, and intermediate strategies that recombine or restructure existing sequences, analogous to chromothripsis \cite{stephens2011massive} or EvoAug-style perturbations, offer a compromise at the cost of added experimental complexity.

Although closed-loop, experiment-in-the-loop pipelines are gaining momentum \cite{morrow2024ml,friedman2025active,crnjar2025pioneer,yin2025iterative,ribeiro2025iterative,castillo2025programming}, the field still lacks systematic benchmarks. Closing this gap will require tighter integration between modeling and experimental design, with candidate sequence pools, acquisition functions, and experimental constraints considered jointly. More broadly, this perspective reframes experimental design from generating data to answer predefined biological questions toward generating data that most effectively improve AI models, using model-identified gaps to guide which measurements are most informative for advancing generalization and, ultimately, discovery.

\section*{Generalization Across Cellular Contexts} 

Seq2func models predict static snapshots of regulatory activity solely from DNA sequence and therefore implicitly assume a fixed cellular context. Chromatin state and \textit{trans}-acting factors, including TFs, cofactors, chromatin regulators, and their abundances, are treated as constant. In reality, regulatory output depends jointly on sequence and dynamic cellular states, so the same sequence can produce different outcomes across cell types, developmental stages, or perturbation conditions. From the model's perspective, cellular context thus acts as an unobserved confounder: it influences regulatory output but is not provided as an input. When models are applied to new cellular contexts, this omission manifests as a concept shift, in which the sequence–function mapping itself changes despite identical inputs.

A common response is scale: multitask training across many cell types to expose models to broader cellular variation. While this improves coverage across cellular diversity, generalization remains limited when test contexts differ substantially from those seen during training. A complementary strategy supplies auxiliary measurements of cellular context, such as chromatin accessibility or histone modification profiles, alongside DNA sequence \cite{karbalayghareh2022chromatin,zhang2023generalizable,lin2024epinformer,murphy2024predicting,javed2025multi,fu2025foundation}. Although this approach can improve performance when such measurements are available, it ties generalization to assay availability and measurement quality. One attempt to bridge this dependence is cross-assay or multimodal imputation, in which missing cellular measurements are inferred from available assays \cite{schreiber2020avocado,nair2025nona,cui2025towards}.

A more flexible alternative is to conditions seq2func models on low-dimensional embeddings of cellular state learned by cell-atlas foundation models \cite{szalata2024transformers,dimitrov2026interpretation}. These embeddings provide contextual signals that can modulate internal representations and support learning of cell-type-specific regulatory programs \cite{nair2019integrating,hingerl2025scooby,lal2024decoding,lin2024epinformer,miao2025xchrom,aksu2025context}. A central challenge is how to integrate such embeddings effectively. Existing approaches range from simple feature concatenation to explicit conditioning mechanisms, including hypernetworks \cite{ha2016hypernetworks} and feature-wise linear modulation (FiLM) \cite{perez2018film}. The goal is to allow shared sequence features to be interpreted differently across cellular contexts by modulating parameters or information flow. However, only a narrow portion of this design space has been explored. Alternatives conditioning strategies, such as channel-wise gating \cite{hu2018squeeze}, low-rank adaptation \cite{hu2022lora}, and routing-based strategies, including mixture-of-experts architectures \cite{shazeer2017outrageously}, remain largely unexplored in regulatory genomics.

Despite their promise, cell-aware seq2func approaches face fundamental limits. Generalization to new cellular contexts is constrained by the diversity of regulatory programs represented during training. When held-out contexts depend on motifs,  chromatin configurations, or regulatory interactions absent from the training data, conditioning alone is insufficient: models must acquire new regulatory rules rather than reweight existing ones. Performance further depends on the fidelity of cellular embeddings. Current single-cell foundation models capture cell identity but often fail to encode cellular dynamics and perturbation responses \cite{ahlmann2025deep,vinas2025systema}, limiting their ability to modulate \textit{cis}-regulatory mechanisms under changing conditions.

Together, these considerations suggest that robust generalization across cellular contexts requires more than appending cell-state features as auxiliary inputs. Progress will depend on tighter coupling between seq2func models, which map sequence to expression, and transcriptomic foundation models that infer cellular state from expression profiles. One promising direction is a bidirectional interface in which seq2func models predict sequence perturbation effects conditioned on cellular embeddings, transcriptomic models propagate those effects through their learned representation of cell state, and the updated state is fed back to refine subsequent sequence predictions. This bidirectional exchange supports joint reasoning about how local \textit{cis}-regulatory changes scale into cell-wide responses and reshape regulatory behavior. Designing and training such integrated systems remains an open challenge. Closer alignment between \textit{cis}- and \textit{trans}-regulatory modeling communities will be essential for developing unified, mechanistically grounded virtual-cell models \cite{bunne2024build}.

\begin{figure}[!t]
     \centering
     \includegraphics[width=0.9\linewidth]{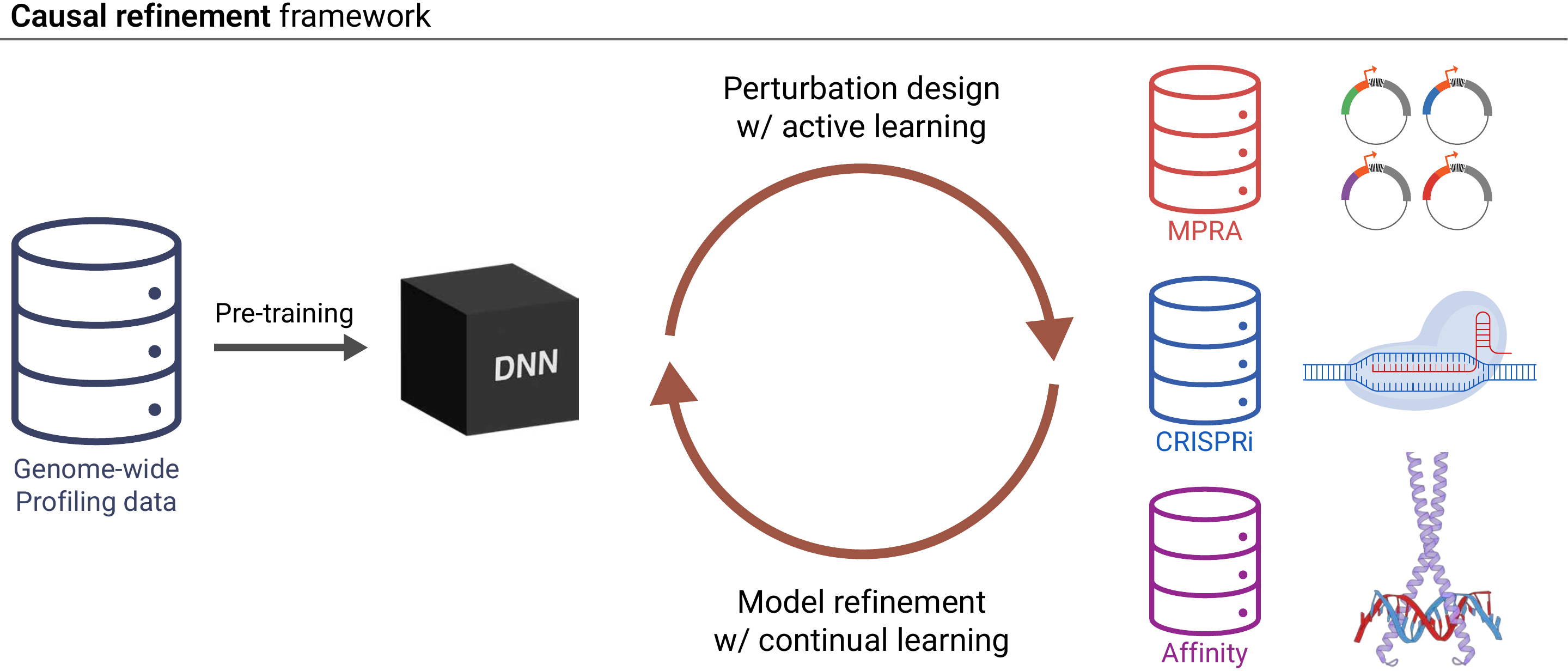}
     \caption{Overview of causal refinement framework. A seq2func model is first pre-trained on large corpus of genome-wide profiling data to learn broad regulatory patterns across cellular contexts. The model then enters phase two, an iterative loop in which targeted perturbation assays, such as MPRA and CRISPRi, are designed using active learning strategies. These experiments generate new measurements that are integrated back into the model through continual learning. Each iteration incrementally refines the model's understanding of \textit{cis}-regulatory mechanisms one locus at a time, progressively improving broader regulatory knowledge across the genome.}
    
     \label{fig:causal_refinement}
 \end{figure}

\section*{Future Outlook: Self-Improving Genomic AI} 

Genomic AI has largely followed a static paradigm, in which models are trained once on observational data and evaluated without structured mechanisms for improvement. Even as seq2func models have scaled across assays and cellular contexts, failures under perturbation, engineered sequences, or novel cellular states are typically analyzed retrospectively rather than used to refine the models.

A natural path beyond this static workflow is what we propose as \emph{causal refinement} (Fig.~\ref{fig:causal_refinement}): a closed-loop process in which a genome-wide model identifies loci or contexts where predictions fail; active learning prioritizes targeted perturbation experiments that expose missing or misrepresented regulatory mechanisms; and continual learning integrates the resulting measurements while preserving genome-wide performance. Iterated over time, this loop accumulates local mechanistic insight that progressively generalizes across the genome. This framework naturally accommodates diverse experimental modalities, including biophysical measurements of TF binding, splicing regulation, and transcriptional dynamics. 

A practical starting point is to establish causal refinement in human cell lines, where genome-wide profiling and high-throughput perturbation screens can be generated efficiently under controlled conditions. Extending this framework to primary cells, tissues, developmental trajectories, and disease models presents a greater challenge, as perturbation data are scarcer and more costly to deploy. In such settings, genome-wide assays such as ATAC-seq and RNA-seq can establish a baseline regulatory landscape, after which targeted perturbations can convert context-specific correlations into mechanistic learning. As \textit{cis}-regulatory knowledge accumulates, progressively smaller perturbation campaigns should suffice to adapt models to new systems.

Similar limitations arise beyond seq2func modeling. Despite training on millions of cells, single-cell transcriptomic foundation models also rely primarily on observational data and consequently exhibit weak generalization under perturbation \cite{ahlmann2025deep,vinas2025systema}. Their failure modes mirror those of seq2func models, reinforcing that scaling observational data alone is insufficient and that progress in both domains requires refinement grounded in perturbation evidence.

Realizing self-improving genomic AI will require perturbation data across many biological systems, motivating a distributed, community-driven effort. In this ecosystem, experimental groups contribute datasets from systems of interest and, in return, gain access to increasingly capable models calibrated to their biological contexts. As predictive gaps narrow, seq2func models can begin to function as virtual instruments, enabling systematic \textit{in silico} perturbation experiments at scales inaccessible to the wet lab and supporting exploration of sequence space, regulatory logic, and counterfactual scenarios.

For this vision to be broadly useful, seq2func models must operate as accessible scientific tools usable by experimental biologists, geneticists, and clinicians, including researchers without extensive computational expertise. This requires user-facing interfaces for querying, interpretation, and experiment design, together with architectures that support fast, memory-efficient inference. Emphasis therefore shifts away from continued scaling of model size and toward compact, deployable models enabled by knowledge distillation and GPU-optimized execution \cite{zhou2026uncertainty,hingerl2025flashzoi,jung2025enigma}. Emerging directions include intuitive chat-based interfaces for natural-language querying \cite{de2025multimodal,schaefer2025multimodal}, automated analysis pipelines \cite{huang2025biomni}, and agentic AI systems for experiment design and execution \cite{swanson2025virtuallab}.

Together, this review positions seq2func models as a foundational framework for regulatory genomics, integrating genome-wide assays, targeted perturbations, and biophysical measurements to connect DNA sequence with regulatory mechanisms. Causal refinement provides a practical path toward self-improving models whose behavior is progressively aligned with causal regulatory rules. Seq2func models in turn complement virtual-cell initiatives that model transcriptional programs and cellular states downstream of gene expression \cite{bunne2024build}. Integrating these approaches creates a domain-specific world model of gene regulation, linking sequence, regulation, expression, and cellular phenotype, and enabling structured, counterfactual reasoning about how sequence changes alter regulatory activity. In this framing, the goal is not a full digital twin of the cell, but a mechanistically grounded framework that iteratively improves through targeted perturbation experiments and continual learning, accelerating discovery and enabling improved predictive models of human health and disease.


\section*{Acknowledgements}\label{sec6}
This work was supported in part by the Simons Center for Quantitative Biology at Cold Spring Harbor Laboratory and NIH grants R01HG012131 and R01GM149921. 

\section*{Competing interests}\label{sec8}
The authors declare no competing interests.

\bibliography{references}
\end{document}